\documentclass[%
reprint,
superscriptaddress,
 amsmath,amssymb,
 aps,
floatfix,
]{revtex4-1}
\usepackage{subfigure}
\usepackage{graphicx}
\usepackage{dcolumn}
\usepackage{bm}
\usepackage[super]{nth}
\usepackage{lipsum}

\begin{document}

\title{Anisotropic magnetic interactions in hexagonal AB-stacked kagome lattice structures: Applications to $\mathrm{Mn}_3\mathrm{X}$ ($\mathrm{X}$ = $\mathrm{Ge}$, $\mathrm{Sn}$, $\mathrm{Ga}$) compounds}

\author{A. Zelenskiy}
 \affiliation{Department of Physics and Atmospheric Science, Dalhousie University, Halifax, Nova Scotia, Canada B3H 3J5}

\author{T. L. Monchesky}
\affiliation{Department of Physics and Atmospheric Science, Dalhousie University, Halifax, Nova Scotia, Canada B3H 3J5}%

\author{M. L. Plumer}
 \affiliation{Department of Physics and Atmospheric Science, Dalhousie University, Halifax, Nova Scotia, Canada B3H 3J5}
 \affiliation{Department of Physics and Physical Oceanography, Memorial University of Newfoundland, St. John’s, Newfoundland, A1B 3X7, Canada}
 
\author{B. W. Southern}
 \affiliation{Department of Physics and Astronomy, University of Manitoba, Winnipeg, Manitoba, Canada R3T 2N2}

\date{\today}

\begin{abstract}
$\mathrm{Mn}_3\mathrm{X}$ compounds in which the magnetic $\mathrm{Mn}$ atoms form AB-stacked kagome lattices have received a tremendous amount of attention since the observation of the anomalous Hall effect in $\mathrm{Mn}_3\mathrm{Ge}$ and $\mathrm{Mn}_3\mathrm{Sn}$. 
Although the magnetic ground state has been known for some time to be an inverse triangular structure with an induced in-plane magnetic moment, there have been several controversies about the minimal magnetic Hamiltonian.
We present a general symmetry-based model for these compounds that includes a previously unreported interplane Dzyaloshinskii-Moriya interaction, as well as anisotropic exchange interactions. 
The latter are shown to compete with the single-ion anisotropy which strongly affects the ground state configurations and elementary spin-wave excitations. 
Finally, we present the calculated elastic and inelastic neutron scattering intensities and point to experimental assessment of the types of magnetic anisotropy in these compounds that may be important.
\end{abstract}

\pacs{Valid PACS appear here}
\maketitle

\section{\label{sec:intro}Introduction}

Kagome lattice antiferromagnets have been the center of attention in many branches of condensed matter physics due to their rich electronic and magnetic properties.
In the quantum limit, these materials are believed to provide a promising platform for experimental realisation of quantum spin liquids and other unconventional phases~\cite{Fujihala2020,Hirschberger2019,Jiang_2019,MENDELS2016455,PhysRevB.87.060405,PhysRevLett.109.067201,PhysRevLett.97.147202,Wen2019,Yan2011}.
On the other hand, the interest in semi-classical non-collinear magnets on kagome lattices has been renewed by recent studies of their interactions with electric currents.
In particular, very recently a few hexagonal D$0_{19}$ compounds with general formula $\mathrm{Mn}_3\mathrm{X}$, ($\mathrm{X} = \mathrm{Ge}, \mathrm{Sn}, \mathrm{Ga}$) were predicted and then experimentally shown to display a large anomalous Hall effect (AHE)~\cite{Chen_PhysRevLett.112.017205,Nakatsuji2015,Nayake1501870,Kiyohara_PhysRevApplied} and topological Hall effect~\cite{Yan}.
AHE in ferromagnetic materials has been studied extensively over the years~\cite{Nagaosa_RevModPhys.82.1539}.
However, more recently it was discovered that it depends not only on the broken time-reversal symmetry but also on the particular type of the magnetic order and the underlying magnetic interactions.
Thus, for example, it has been well established that in compounds with collinear ferromagnetic order, spin-orbit coupling is crucial for the AHE~\cite{Berger_PhysRevB.2.4559}. 
Unlike collinear antiferromagnets, non-collinear antiferromagnetic structures have been shown to induce AHE even without spin-orbit coupling.
However, little is known about the types of antiferromagnetic order that can yield AHE and the majority of the existing theories have been based on the previous studies of $\mathrm{Mn}_3\mathrm{Sn}$ and $\mathrm{Mn}_3\mathrm{Ge}$~\cite{Chen_PhysRevLett.112.017205,K_bler_2014,Busch_PhysRevResearch.2.033112}.
From a technological standpoint these emergent transport properties of $\mathrm{Mn}_3\mathrm{X}$ compounds are very attractive in the development of antiferromagnetic spintronics and memory devices since the size of these materials is not limited by the demagnetizing fields as in the case of ferromagnets.

Novel techniques have recently been proposed for imaging and writing of magnetic domains in $\mathrm{Mn}_3\mathrm{Sn}$~\cite{Reichlova2019}.
\begin{figure}[!t]
    \centering
    \includegraphics[width=0.47\textwidth]{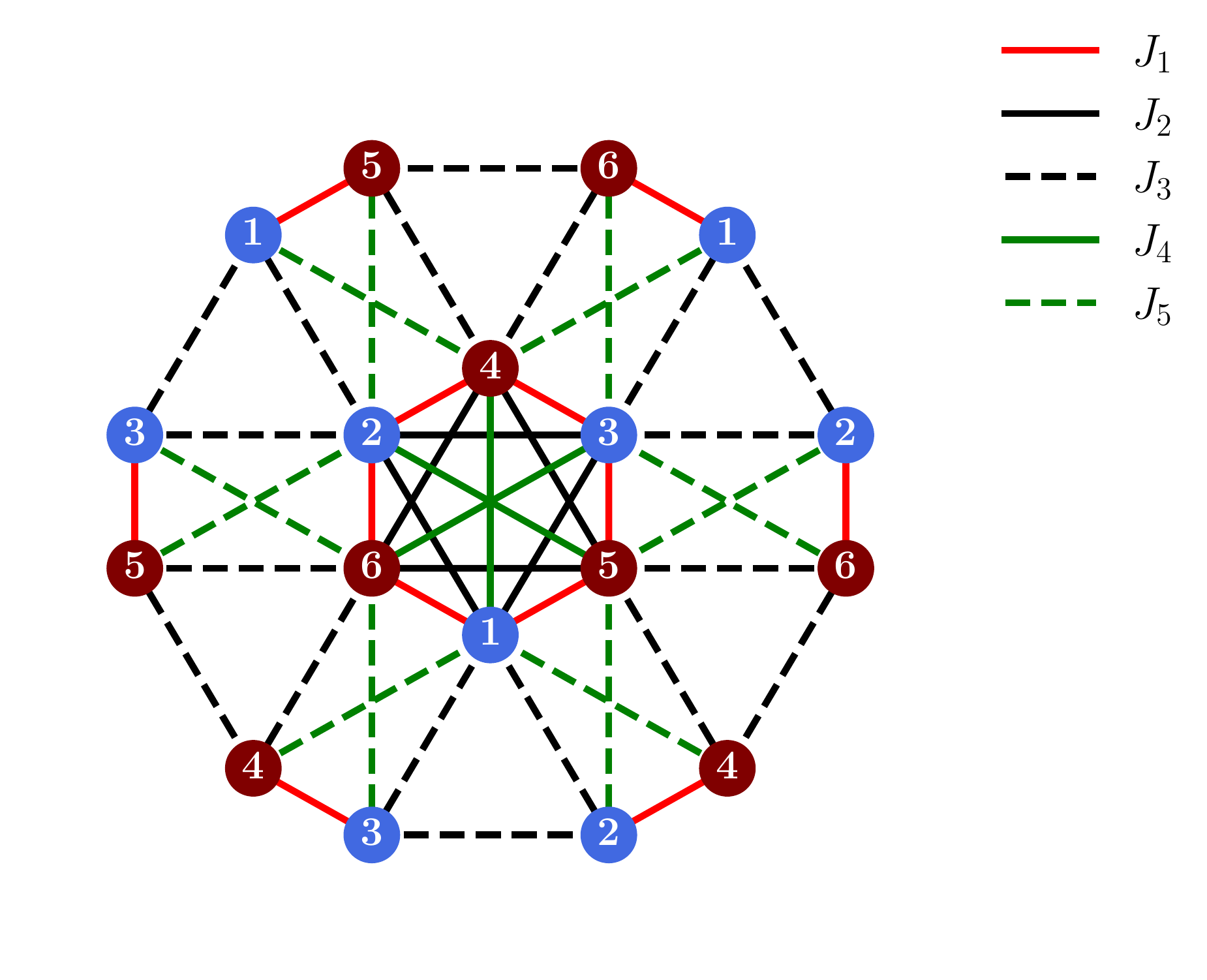}
    \caption{Magnetic exchange pathways in the $\mathrm{Mn}_3\mathrm{X}$ AB-stacked kagome crystals. Here, the circles represent the $\mathrm{Mn}$ atoms, and light blue and dark red indicate atoms with $z=\frac{1}{4}$ and $z=\frac{3}{4}$ respectively. The numbers further label the six sublattices.}
    \label{fig:Interactions}
\end{figure}
\noindent
Consequently, there have been several experimental and theoretical studies focused on determination of the magnetic ground state of $\mathrm{Mn}_3\mathrm{X}$ compounds.
Yasukochi~\cite{Yasukoshi} and Ohoyama~\cite{Ohoyama} were the first to identify weak ferromagnetism in $\mathrm{Mn}_3\mathrm{Sn}$ and $\mathrm{Mn}_3\mathrm{Ge}$.
Later, the first neutron diffraction studies of these two compounds led to determination of the non-collinear 120$^\circ$ structure~\cite{kouvel1965proceedings}.
Consecutive powder neutron diffraction experiments on  $\mathrm{Mn}_3\mathrm{Ga}$~\cite{KREN19701653}, $\mathrm{Mn}_3\mathrm{Ge}$~\cite{kadar1971neutron}, and $\mathrm{Mn}_3\mathrm{Sn}$~\cite{Zimmer} also determined that the antiferromagnetic order and the induced magnetic moment are restricted to the plane perpendicular to the $\hat{c}$-axis of the compound.
More recent studies revealed that of the possible triangular magnetic structures, the ground state of these $\mathrm{Mn}_3\mathrm{X}$ compounds is the antichiral ``inverse triangular'' structure~\cite{Brown_1990,Soh_PhysRevB,*Soh_supp,Chen_PhysRevB.102.054403,Duan}.

The main challenge in the modelling of the magnetic properties of these compounds is the abundance of magnetic interactions, as evident from Fig.~\ref{fig:Interactions}.
The lattice structure, which consists of corner-sharing equilateral triangles, establishes competing antiferromagnetic exchange interactions leading to geometric frustration.
The exchange couplings originate predominantly from the Ruderman-Kittel-Kasuya-Yosida (RKKY) interactions which is characteristic of the metallic magnetic materials with itinerant $d$-electrons~\cite{Park2018}.
Unlike regular kagome lattices, the crystal structure of $\mathrm{Mn}_3\mathrm{X}$ family is referred to as \textit{breathing} kagome as it contains adjacent triangles of slightly different size.
This results in anisotropy in the exchange interactions between the spins belonging to different triangles.
From the values of the crystallographic parameters~\cite{Chen_PhysRevB.102.054403,Park2018,Brown_1990,Khmelevskyi_2016,CHANGTAOFAN1966}, the differences in bond lengths for the two types of triangles in $\mathrm{Mn}_3\mathrm{Ge}$, $\mathrm{Mn}_3\mathrm{Ga}$ and $\mathrm{Mn}_3\mathrm{Sn}$ are approximately $0.008$ \AA, $0.08$ \AA, and $0.2$ {\AA} respectively.
As a result, the breathing anisotropy is expected to be largest in $\mathrm{Mn}_3\mathrm{Sn}$ and nearly negligible in the other two compounds.
In the present work, the effects of this type of anisotropy will be omitted.
All previous studies included both in-plane ($J_2$ and $J_3$) and out-of-plane ($J_1$) nearest-neighbour (NN) exchange interactions; however, some studies also indicate the importance of the next-nearest-neighbour (NNN) exchange interactions~\cite{Cable_PhysRevB.48.6159,Park2018,Chen_PhysRevB.102.054403}.
Most studies also included in-plane NN Dzyaloshinskii-Moriya (DM) interactions, although the form of the DM vector has been inconsistent in some of the recent literature: In most cases the DM vector is chosen to be perpendicular to the kagome planes, $\mathbf{D_{ij}}\parallel \mathbf{\hat{z}}$. However, Ref.~\cite{Liu_PhysRevLett.119.087202} also includes a term with a DM vector along the in-plane bond directions.

Lastly, there is an ongoing controversy about the role of the magnetocrystalline anisotropy in these compounds.
It is widely known that due to geometric frustration, the ground state of a 2D kagome antiferromagnet with NN interactions only is a 120$^\circ$ structure with a macroscopic $U(1)$ degeneracy.
Early inelastic neutron scattering experiments~\cite{Radhakrishna_1991,Cable_PhysRevB.48.6159} revealed that the excitation spectrum contains an anisotropy gap that is associated with the in-plane spin fluctuations.
In order to produce this energy gap, some of the previous studies~\cite{Cable_PhysRevB.48.6159,Park2018,Soh_PhysRevB,*Soh_supp} included \nth{6} order single-ion anisotropy since the \nth{2} and \nth{4} order terms cannot break the continuous manifold of the 120$^\circ$ ground state configuration~\cite{Cable_PhysRevB.48.6159,Kiyohara_PhysRevApplied,NAGAMIYA1982385,Liu_PhysRevLett.119.087202}.
Nevertheless, it has been reported that due to the deviations from the 120$^\circ$ structure induced by weak ferromagnetism in $\mathrm{Mn}_3\mathrm{X}$ compounds, the previous arguments no longer apply and \nth{2} order anisotropy is expected to be sufficient to break the $U(1)$ degeneracy~\cite{Chen_PhysRevB.102.054403}.
This observation is relevant to the present study.

In this paper, we present a general magnetic Hamiltonian model for the AB-stacked hexagonal family of compounds, derived from symmetry considerations in hopes of resolving some of the existing controversies about the magnetic interactions in these $\mathrm{Mn}_3\mathrm{X}$ compounds.
This model is then used to investigate the relative effects of the single-ion and the exchange anisotropy on the magnetic structure of the ground state spin configurations, and in particular on the induced in-plane magnetic moment.
Based on these results, we provide calculations of the elastic neutron scattering intensities for the systems with different types of magnetic anisotropy.
The impact of these anisotropies on spin waves and inelastic neutron scattering intensities is then examined.

\section{\label{sec:model}Model}

\subsection{\label{subsec:structure} Structural details}

Hexagonal D$0_{19}$ compounds belong to P6$_3/mmc$ (No. 194) space group.
In the case of the $\mathrm{Mn}_3\mathrm{X}$ family, the six $\mathrm{Mn}$ atoms are located at the $6h$ Wyckoff positions and form the AB-stacked breathing kagome lattice planes, while the non-magnetic $\mathrm{X}$ atoms sit in the centers of the $\mathrm{Mn}$ hexagons (Wyckoff position $2c$).
The atomic coordinates of the six $\mathrm{Mn}$ atoms are $\left(x,2x,\frac{1}{4}\right)$, $\left(x,\bar{x},\frac{1}{4}\right)$, $\left(2\bar{x},\bar{x},\frac{1}{4}\right)$, $\left(\bar{x},2\bar{x},\frac{3}{4}\right)$, $\left(\bar{x},x,\frac{3}{4}\right)$, $\left(2x,x,\frac{3}{4}\right)$ where $x$ determines the breathing amplitude of the lattice. 
When $x=\frac{5}{6}$, the structure simplifies to a perfect kagome lattice.

\subsection{\label{subsec:model} Magnetic Hamiltonian}

\begin{figure}[!t]
    \centering
    \includegraphics[width=0.35\textwidth]{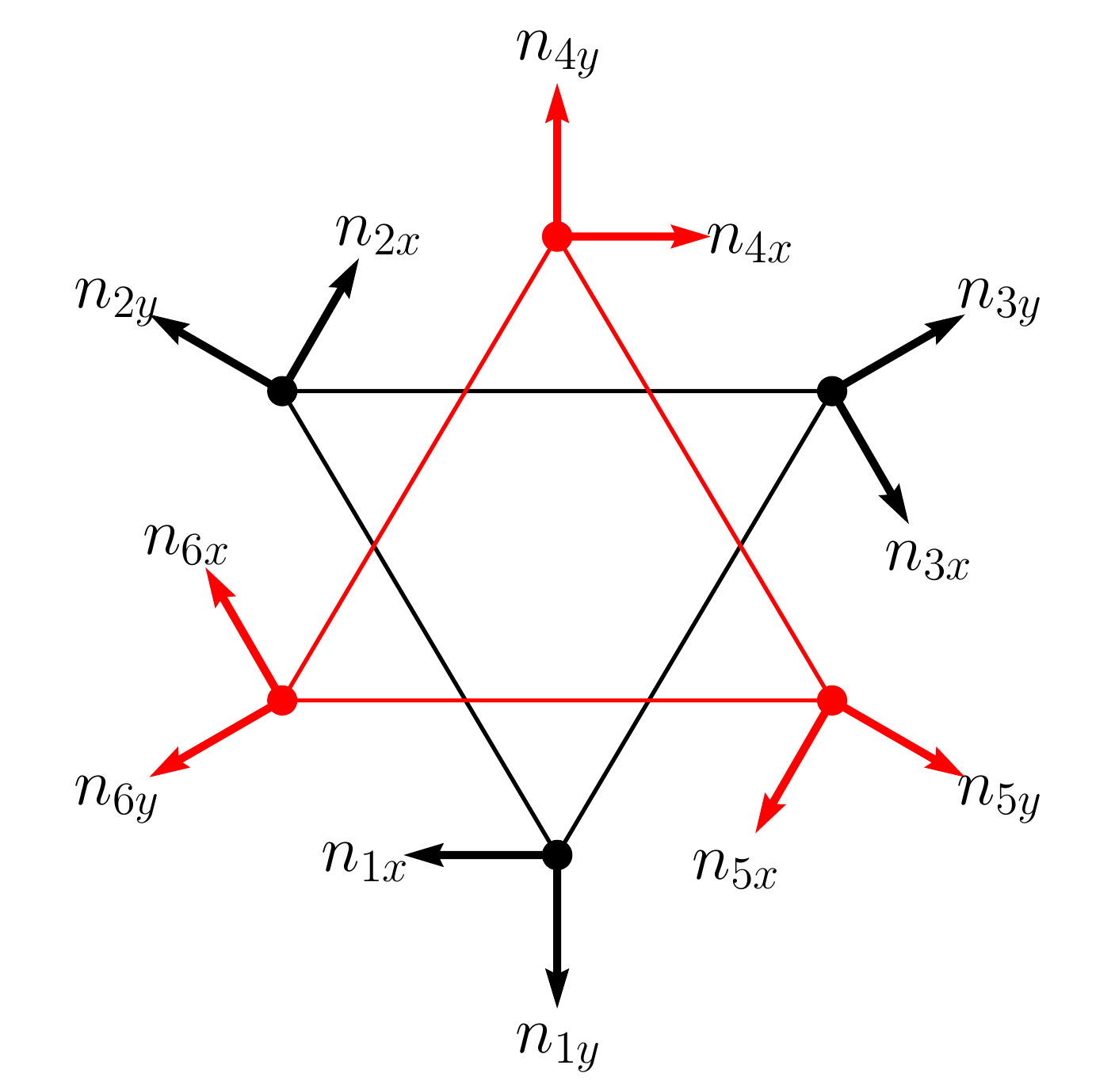}
    \caption{Local anisotropy axes for the six sublattices of $\mathrm{Mn}_3\mathrm{X}$. The local $z$-axes have the same direction (out of the page). The full vector expressions are given in Supplemental Material.}
    \label{fig:local axes}
\end{figure}

Since the overall spin energy must be invariant under all symmetry transformations of the space group of the crystal system, the magnetic model is constructed by identifying all of the spin invariants.
For the purpose of this paper, only terms quadratic in spin components were considered.
The full derivation of the model is provided in the Supplemental Material.
The corresponding spin Hamiltonian is given by

\begin{align}
    &\mathcal{H} = \mathcal{H}_{K} + \mathcal{H}_{J} + \mathcal{H}_{D} + \mathcal{H}_{A} \label{eq: magnetic_hamiltonian}\\
    &\mathcal{H}_{K} = \sum_\mathbf{r}\sum_{i}\sum_\alpha K_{\alpha} \left(\mathbf{\hat{n}}_{i\alpha}\cdot \mathbf{S}_{i}(\mathbf{r})\right)^2\notag\\
    &\mathcal{H}_{J} = \sum_{\mathbf{r}\mathbf{r'}}\sum_{ij} J_{ij}(\mathbf{r}-\mathbf{r}') \mathbf{S}_{i}(\mathbf{r})\cdot\mathbf{S}_{j}(\mathbf{r'})\notag\\
    &\mathcal{H}_{D} = \sum_{\mathbf{r}\mathbf{r'}}\sum_{ij} D_{ij}(\mathbf{r}-\mathbf{r}')\mathbf{\hat{z}}\cdot \left(\mathbf{S}_{i}(\mathbf{r})\times\mathbf{S}_{j}(\mathbf{r'})\right)\notag\\
    &\mathcal{H}_{A} = \sum_{\mathbf{r}\mathbf{r'}}\sum_{ij}\sum_\alpha A_{ij\alpha}(\mathbf{r}-\mathbf{r'})\left(\mathbf{n}_{i\alpha}\cdot\mathbf{S}_{i}(\mathbf{r})\right)\left(\mathbf{n}_{j\alpha}\cdot\mathbf{S}_{j}(\mathbf{r'})\right),\notag
\end{align} where $\mathcal{H}_{K}$ is the \nth{2} order single-ion anisotropy, $\mathcal{H}_{J}$ is the isotropic Heisenberg exchange, $\mathcal{H}_{D}$ is the DM interaction, and $\mathcal{H}_{A}$ is the symmetric, anisotropic exchange interaction.
The latter interactions have also been derived for two-dimensional kagome planes~\cite{Essafi_PhysRevB.96.205126}.
However, the interplane exchange anisotropies have not been reported before.
Sum indices $\mathbf{r}$, $\mathbf{r'}$ label unit cells, $i, j\in\{1,...,6\}$ label atoms in each unit cell, and $\alpha \in \{x,y,z\}$ labels the spin vector components.  
Vectors $\mathbf{n}_{i\alpha}$ represent local anisotropy axes and are shown in Fig.~\ref{fig:local axes}.
\noindent
In total, we identify three \nth{2} order single-ion anisotropy terms with anisotropy constants $K_x$, $K_y$, and $K_z$, five isotropic exchange interactions with coupling constants $J_1-J_5$, three DM interaction terms with DM vectors $\mathbf{D}_1$, $\mathbf{D}_2$, and $\mathbf{D}_3$ and ten anisotropic exchange interactions with coupling constants $A_i = A_{ix} = -A_{iy}$, $A_{iz}$ with $i\in{1,...,5}$.
Note that in the case of single-ion anisotropy, there are only two independent coupling constants since the magnitude of the spins is taken to be fixed. 
The notation used throughout this paper is chosen based on the distance between the magnetic ions: index 1 labels out-of-plane NN interactions, 2 and 3 label in-plane NN interactions, and 4,5 label NNN interactions. 
Note also that the symmetry of the lattice restricts all of the DM vectors to point perpendicular to the kagome planes.
As mentioned in the Introduction, the in-plane DM interactions have already been implemented in some of the previous studies, however, to our knowledge, the out-of-plane DM interaction has not been considered before.
Similarly, the exchange anisotropy has not been used in any of the previous studies of $\mathrm{Mn}_3\mathrm{X}$ systems.

The exchange anisotropy, also called bond-dependent anisotropy, typically originates from the strong spin-orbit coupling~\cite{SOC_exchange_anis}. 
However, unlike DM interactions, the existence of exchange anisotropy does not depend on the inversion symmetry of a crystal.
In magnetic insulators, similar terms have been considered, such as the compass and Kitaev interactions~\cite{Trousselet_2010,Compass_review,Nikolaev_compass}.
In triangular lattices, these interactions have been shown to stabilize spiral and multi-$Q$ spin configurations~\cite{hayami2020noncoplanar}.

\section{\label{sec:ground_state}Magnetic ground state}
\subsection{\label{subsec: single plane}Single layer}
Previous studies have established that the inverse-triangular ground state in two-dimensional kagome systems is stabilized by in-plane DM interaction with a negative DM constant~\cite{Soh_PhysRevB,*Soh_supp,Chen_PhysRevB.102.054403}.
More generally, we find that this state is stable whenever $J_2>0$, $J_3>0$, and $D_2<0$, $D_3<0$.
Therefore, throughout the paper we focus on $\mathrm{Mn}_3\mathrm{Ge}$ as a prototype for hexagonal AB-stacked $\mathrm{Mn}_3\mathrm{X}$ compounds, in which case it is reasonable to set $J_3 = J_2$, $D_3 = D_2$, and $A_3 = A_2$.
The spin Hamiltonian for a single kagome plane simplifies to 
\begin{widetext}
\begin{align}
    \mathcal{H}_p &= K_x\sum_{\mathbf{r}}\Big[ (\mathbf{\hat{n}}_{1x}\cdot \mathbf{S}_1(\mathbf{r}))^2+(\mathbf{\hat{n}}_{2x}\cdot \mathbf{S}_2(\mathbf{r}))^2+(\mathbf{\hat{n}}_{3x}\cdot \mathbf{S}_3(\mathbf{r}))^2\Big] + J_2\sum_{\langle\mathbf{r}\mathbf{r}'\rangle}\Big[\mathbf{S}_1(\mathbf{r})\cdot\mathbf{S}_2(\mathbf{r}')+\mathbf{S}_1(\mathbf{r})\cdot\mathbf{S}_3(\mathbf{r}')+\mathbf{S}_2(\mathbf{r})\cdot\mathbf{S}_3(\mathbf{r}')\Big]\notag\\
    &+ D_2\sum_{\langle\mathbf{r}\mathbf{r}'\rangle}\mathbf{\hat{z}}\cdot\Big[\mathbf{S}_1(\mathbf{r})\times\mathbf{S}_2(\mathbf{r}')-\mathbf{S}_1(\mathbf{r})\times\mathbf{S}_3(\mathbf{r}')+\mathbf{S}_2(\mathbf{r})\times\mathbf{S}_3(\mathbf{r}')\Big]\notag\\
    &+ A_2\sum_{\langle\mathbf{r}\mathbf{r'}\rangle} \Big[\left(\mathbf{n}_{1x}\cdot\mathbf{S}_{1}(\mathbf{r})\right)\left(\mathbf{n}_{2x}\cdot\mathbf{S}_{2}(\mathbf{r'})\right) + \left(\mathbf{n}_{1x}\cdot\mathbf{S}_{1}(\mathbf{r})\right)\left(\mathbf{n}_{3x}\cdot\mathbf{S}_{3}(\mathbf{r'})\right) + \left(\mathbf{n}_{2x}\cdot\mathbf{S}_{2}(\mathbf{r})\right)\left(\mathbf{n}_{3x}\cdot\mathbf{S}_{3}(\mathbf{r'})\right)\Big]\notag\\
    &- A_2\sum_{\langle\mathbf{r}\mathbf{r}'\rangle} \Big[\left(\mathbf{n}_{1y}\cdot\mathbf{S}_{1}(\mathbf{r})\right)\left(\mathbf{n}_{2y}\cdot\mathbf{S}_{2}(\mathbf{r'})\right) + \left(\mathbf{n}_{1y}\cdot\mathbf{S}_{1}(\mathbf{r})\right)\left(\mathbf{n}_{3y}\cdot\mathbf{S}_{3}(\mathbf{r'})\right) + \left(\mathbf{n}_{2y}\cdot\mathbf{S}_{2}(\mathbf{r})\right)\left(\mathbf{n}_{3y}\cdot\mathbf{S}_{3}(\mathbf{r'})\right)\Big],
    \label{eq:single Kagome hamiltonian}
\end{align}
\end{widetext} where $\langle\cdots\rangle$ represents sums over nearest neighbours.
\begin{figure}[!t]
    \centering
    \includegraphics[width=0.48\textwidth]{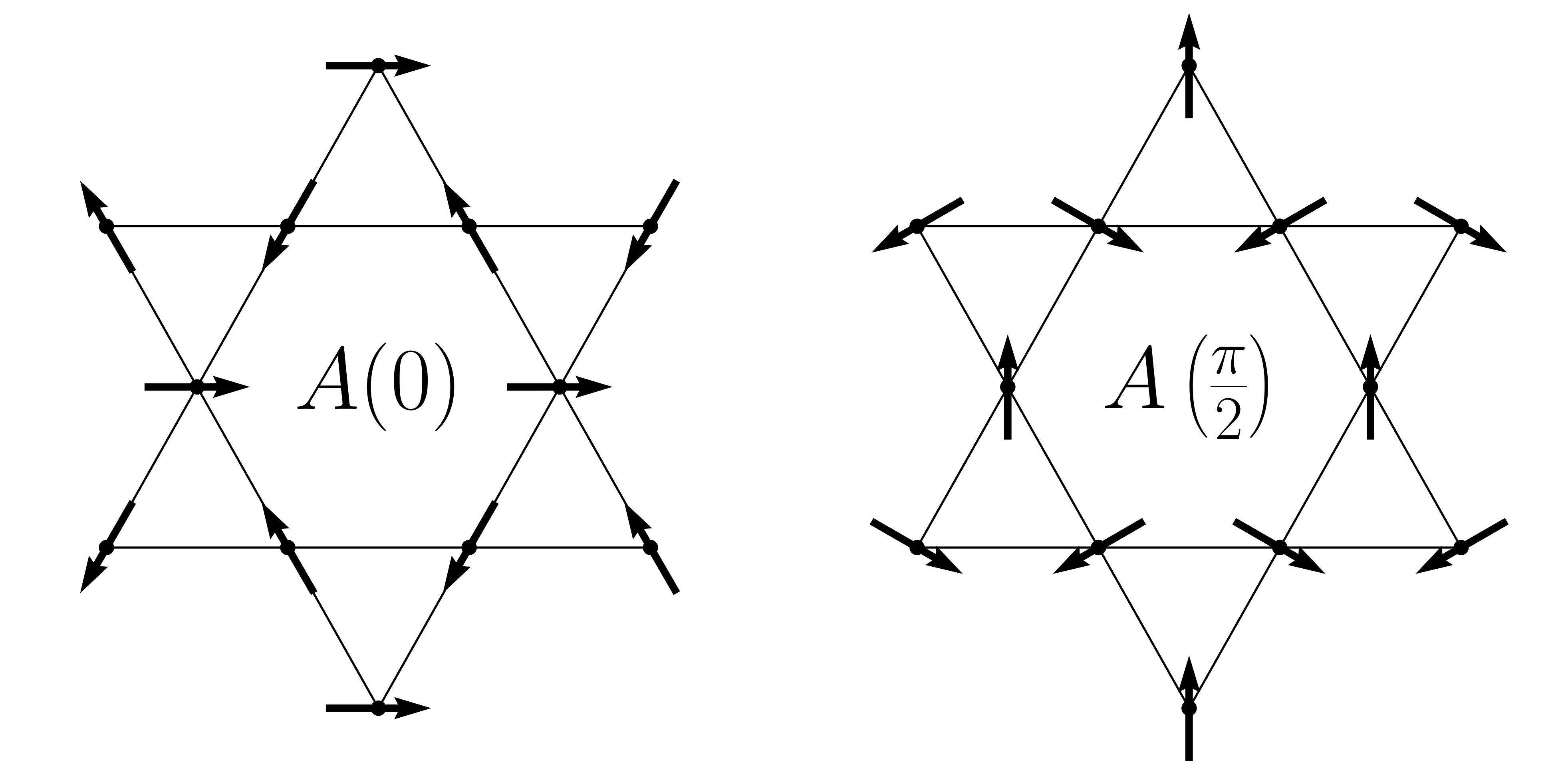}
    \caption{Inverse-triangular 120$^\circ$ structure. The two magnetic configurations shown here are orthogonal and form a two-dimensional order parameter $\mathbf{A}=A(\phi)$ which transforms according to $E_{1g}$ irreducible representation of the point group $D_{6h}$.}
    \label{fig:GS no anis}
\end{figure}
\noindent
The magnetic ground states can be calculated by minimizing the target Hamiltonian using Monte-Carlo simulated annealing~\cite{Annealing}.
The simulations were performed on a system with $6^3$ unit cells with runs at a given temperature consisting of $10^4$ Monte-Carlo steps.
In the case of a continuous ordering phase transition, the ordered phase must transform as one of the irreducible representations of the underlying symmetry group.
The magnetic moments on the magnetic $\mathrm{Mn}$ sites were previously shown to form an 18-dimensional representation~\cite{Soh_PhysRevB,*Soh_supp,Chen2021_nature}.
However, since the experimental results indicate that the ground state spin configurations are planar, and that the spins in the $z=\frac{1}{4}$ sublattice are parallel to their inversion-related partners in the $z=\frac{3}{4}$ sublattice, the representation of the spins can be expressed in a 6-dimensional form.
As outlined in the Supplementary notes of Ref.~\cite{Soh_PhysRevB,*Soh_supp}, this 6-dimensional representation can be decomposed into a combination of three irreducible representations: $B_{1g}\oplus B_{2g}\oplus 2E_{1g}$, where $B_{1g}$ and $B_{2g}$ are 1-dimensional, and $E_{1g}$ is 2-dimensional irreducible representations of the point group $D_{6h}$.
When the anisotropic terms in the Hamiltonian are zero ($K_x =0$, $A_2 = 0$), the ground state spin configuration is the inverse-triangular structure, shown labeled as a two-dimensional vector $\mathbf{A}$ in Fig.~\ref{fig:GS no anis}.
An important feature of the hexagonal $\mathrm{Mn}_3\mathrm{X}$ compounds is that both the in-plane magnetization $\mathbf{M}$ and the order parameter $\mathbf{A}$ transform according to the $E_{1g}$ irreducible representation.
This results in an invariant coupling of the order parameter to the magnetization, which has been previously shown to be $\propto \mathbf{A}\cdot\mathbf{M}$~\cite{Soh_PhysRevB,*Soh_supp}.
As a result, the spin configuration shown in Fig.~\ref{fig:GS no anis} may be distorted by acquiring an in-plane magnetic moment without changing the symmetry of the ground state.
The consequences of the induced magnetic moment are far-reaching. 
When the ground state spin configuration corresponds to order parameter $\mathbf{A}$, the constraint that the spin vectors have fixed magnitudes implies that one can write $\mathbf{A} = A(\phi)$ as function of a single parameter, $\phi$, the angle of the global in-plane spin rotation.
%
\begin{figure}[!t]
    \centering
    \includegraphics[width=0.45\textwidth]{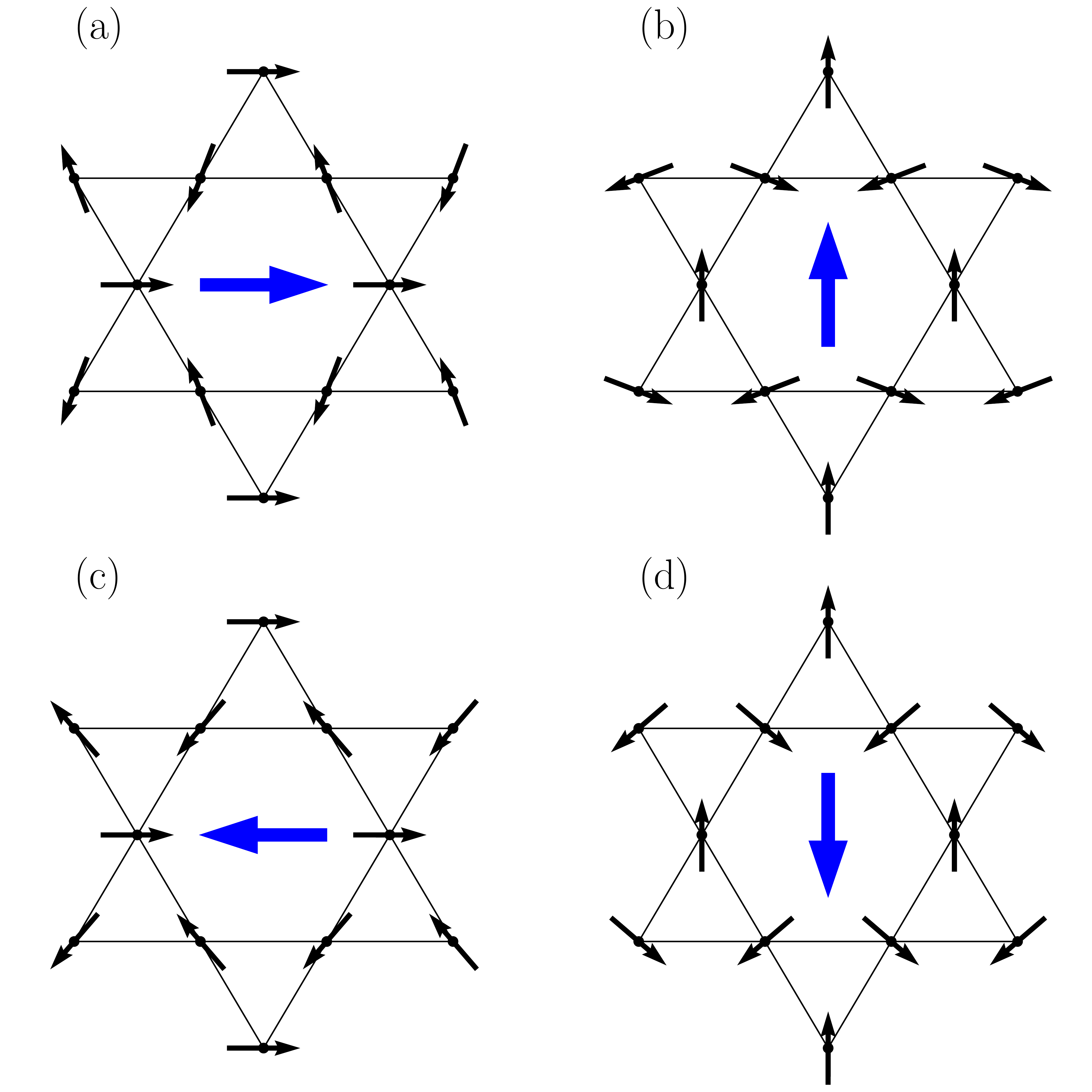}
    \caption{Distorted inverse-triangular structure, $\mathbf{A}+\mathbf{M}$, which results from non-zero single-ion and exchange anisotropy interactions. (a) $K_x<0$, $A_2=0$ (b) $K_x>0$, $A_2=0$, (c) $K_x=0$, $A_2<0$, and (d) $K_x=0$, $A_2>0$. The blue arrow indicates the direction of the induced magnetic moment.}
    \label{fig:GS anis}
\end{figure}
\noindent
When this spin configuration is inserted in (\ref{eq:single Kagome hamiltonian}), one finds that the energy per spin is 
\begin{figure}[h!]
    \centering
    \includegraphics[width=0.48\textwidth]{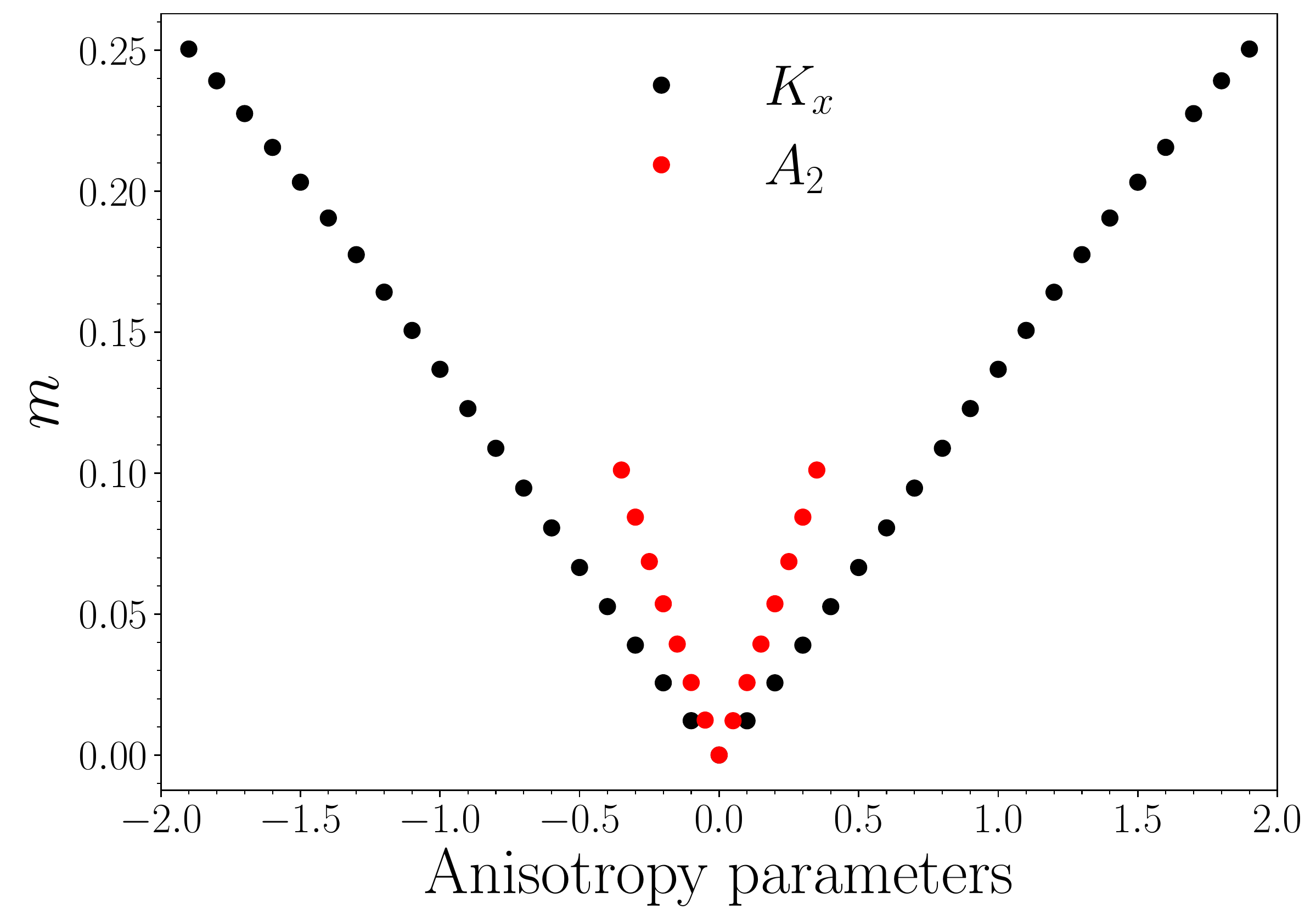}
    \caption{The magnitude of the induced in-plane magnetic moment per spin as function of one of the anisotropy parameters, $K_x$ (black dots) or $A_2$ (red dots), while keeping the other parameter zero.}
    \label{fig: m vs anisotropy}
\end{figure}
\begin{equation}
    E_\mathbf{A} = \sqrt{3}D_2 - J_2 + \frac{K_x}{2},
    \label{eq: energy of A}
\end{equation} which is independent of $\phi$ and $A_2$.
Now, when the 120$^\circ$ configuration is distorted, two spins in the unit cell rotate towards (or away from) each other by a small angle $\theta$, inducing an in-plane magnetic moment.
The magnetic energy can then be written as 

\begin{equation}
    E_{\mathbf{A}+\mathbf{M}} = E_1(J_2, D_2, \theta) + E_2(K_x, A_2,\phi,\theta).
    \label{eq: energy of A+M}
\end{equation} Assuming that the value of the distortion angle is small, these terms can be written as

\begin{align}
    E_1(J_2, D_2, \theta)&\approx (J_2-\sqrt{3}D_2)\theta^2,\\
    E_2(K_x, A_2,\phi,\theta)&\approx\cos{2\phi}\left[\frac{K_x-2A_2}{\sqrt{3}}\theta + \frac{A_2+K_x}{3}\theta^2\right].
    \label{eq: approx energy A+M}
\end{align} 
As a result, the energy has a $\phi$-dependent term which depends on the strength of the anisotropic interactions, $K_x$ and $A_2$.
The minima of $E_{\mathbf{A}+\mathbf{M}}$ are determined by the signs and relative magnitudes of the anisotropic coupling constants as presented in Fig.~\ref{fig:GS anis}.
Moreover, it can be shown that $\theta = 0$ only when both $K_x$ and $A_2$ are zero.
The explicit forms of expressions in (\ref{eq: energy of A+M}) are presented in the Supplemental Material. 
The important result is that the inclusion of anisotropic interactions sets a competition with the in-plane DM interaction that favours the inverse triangular structure.
This, in turn, induces an in-plane magnetic moment, which removes the continuous degeneracy of the $A(\phi)$ configuration associated with the $U(1)$ symmetry.
This is confirmed by the numerical calculations that were also used to determine the magnitude of the induced moment per spin, $m = \frac{|\mathbf{M}|}{N}$, as function of the anisotropy parameters, $K_x$ and $A_2$, where $N$ represents the number of spins. 
Fig.~\ref{fig: m vs anisotropy} shows the results for a system with $J_2 = 1$ and $D_2 = -0.2$.
In the high-anisotropy limit ($|K_x|>2$ or $|A_2|>0.35$), the ground state configuration changes to a perfect 120$^\circ$ structure with spins pointing along the corresponding local anisotropy axes.
In both cases, the relationship between the anisotropic parameters and the magnitude of the magnetic moment is approximately linear, with the slope of the $m$ vs $A_2$ line approximately twice as large as that of the $m$ vs $K_x$ line that is as expected from (\ref{eq: approx energy A+M}).

\subsection{\label{subsec: multiple planes}AB-stacked layers}
When the kagome planes are coupled to each other via the out-of-plane interactions, the Hamiltonian becomes

\begin{figure}[!ht]
    \centering
    \includegraphics[width=0.45\textwidth]{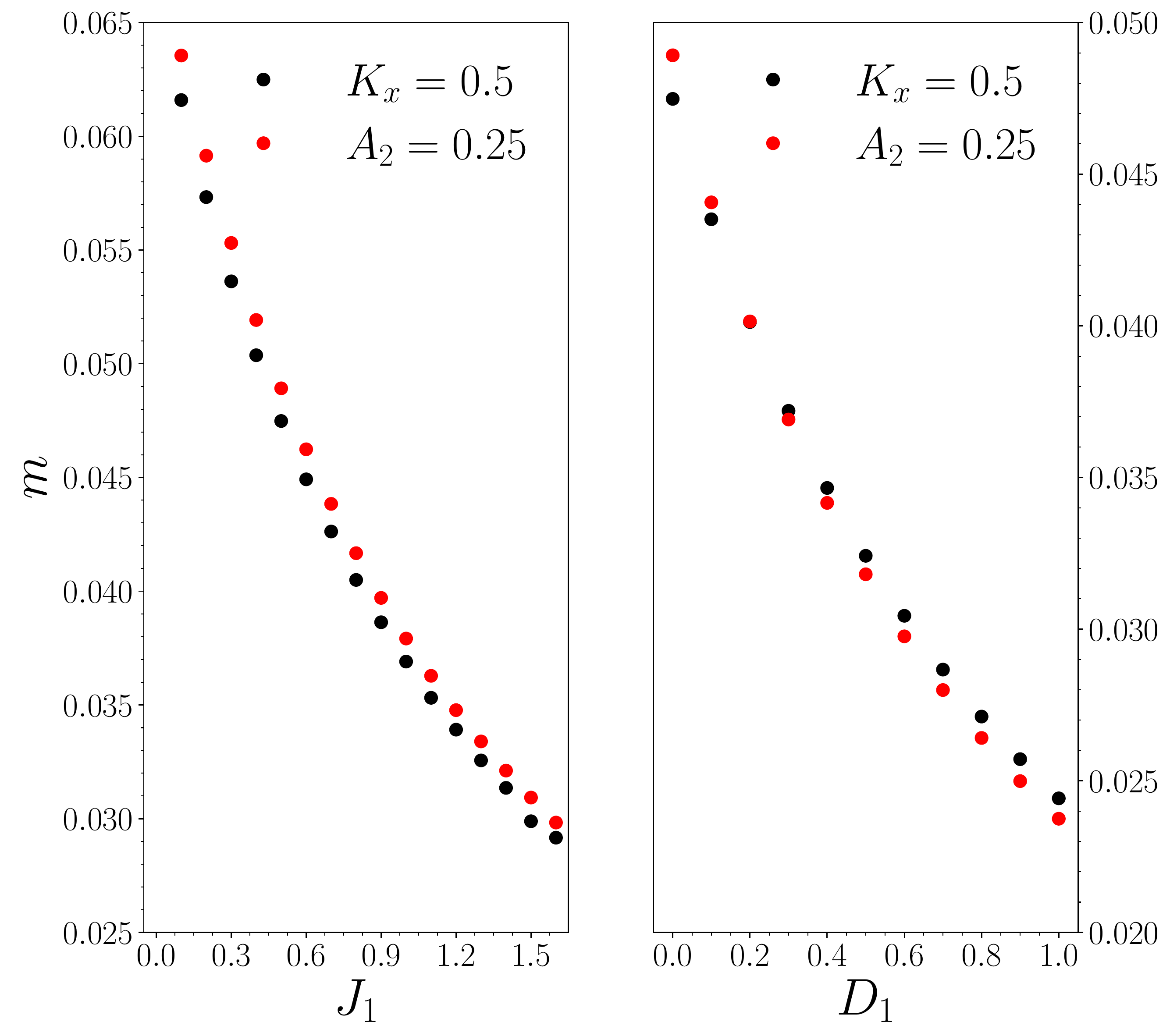}
    \caption{Magnitude of the induced magnetic moment as function of the NN interplane exchange and DM coupling constants. The values of $K_x$ and $A_2$ were chosen to give similar values of $m$ in the decoupled system. When $J_1=0$, the directions of the magnetic moments in different planes are uncorrelated and the net magnetization is zero on average.}
    \label{fig: m vs J1}
\end{figure}

\begin{figure}[!ht]
    \centering
    \includegraphics[width=0.45\textwidth]{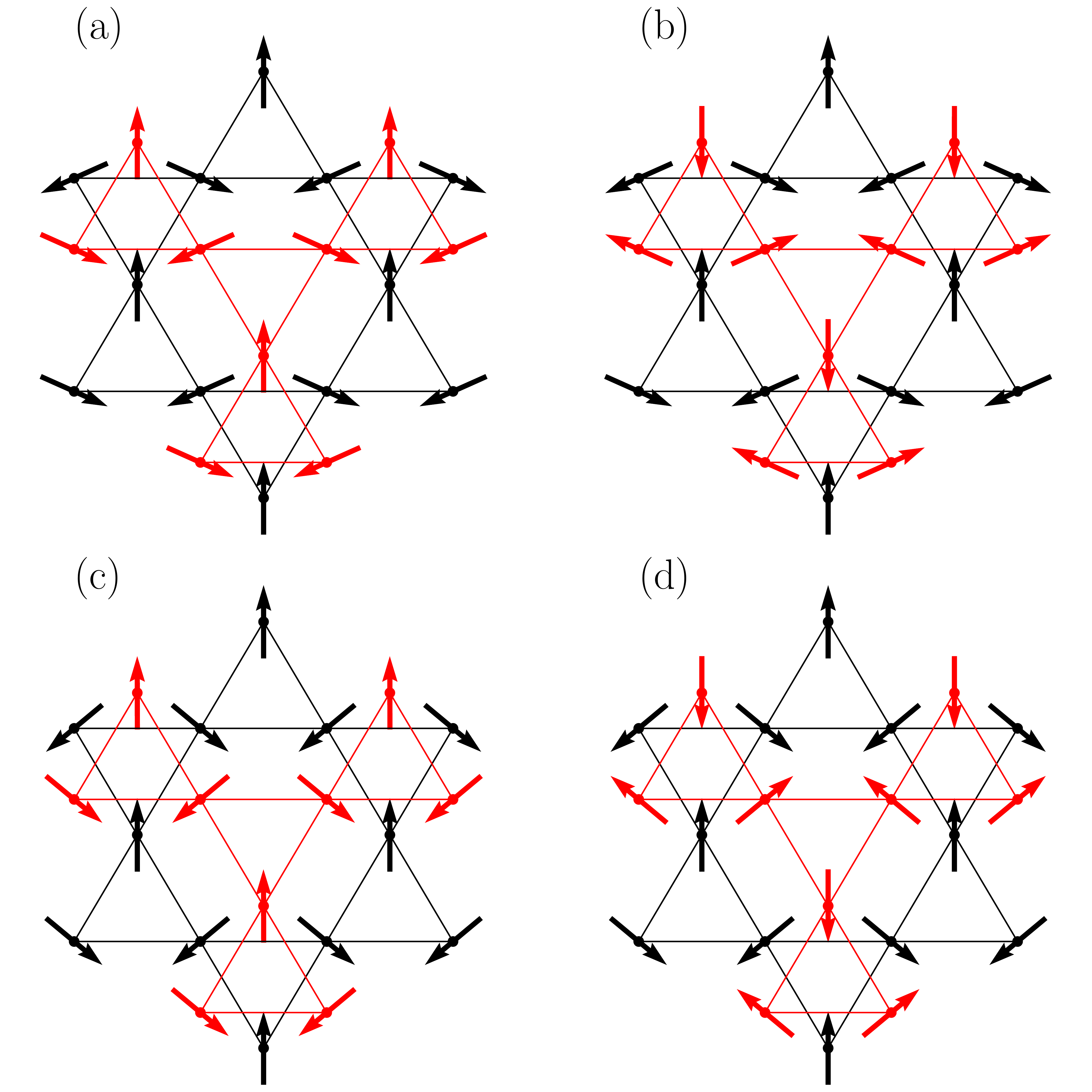}
    \caption{Magnetic ground states for AB-stacked kagome planes coupled via NN out-of-plane exchange interactions ($J_1$). (a) $K_x>0$, $A_2=0$, $J_1>0$, (b) $K_x>0$, $A_2=0$, $J_1<0$, (c) $K_x=0$, $A_2>0$, $J_1>0$, (d) $K_x=0$, $A_2>0$, $J_1<0$. For all cases, two spins in in each triangle rotate towards or away from each other, inducing magnetic moment. Note, however, that the induced in-plane moments cancel in (b) and (d).}
    \label{fig:GS 2 planes anis}
\end{figure}

\begin{widetext}
\begin{align}
    \mathcal{H} &= \sum_{p_1,p_2} \left[\mathcal{H}_{p_1} + \mathcal{H}_{p_2} \right]+ J_1\sum_{\langle\mathbf{r}\mathbf{r}'\rangle}\Big[\mathbf{S}_1(\mathbf{r})\cdot\mathbf{S}_5(\mathbf{r}')+\mathbf{S}_1(\mathbf{r})\cdot\mathbf{S}_6(\mathbf{r}')+\mathbf{S}_2(\mathbf{r})\cdot\mathbf{S}_4(\mathbf{r}')+\mathbf{S}_2(\mathbf{r})\cdot\mathbf{S}_6(\mathbf{r}')+\mathbf{S}_3(\mathbf{r})\cdot\mathbf{S}_4(\mathbf{r}')+\mathbf{S}_3(\mathbf{r})\cdot\mathbf{S}_5(\mathbf{r}')\Big]\notag\\
    &+ D_1\sum_{\langle\mathbf{r}\mathbf{r}'\rangle}\mathbf{\hat{z}}\cdot\Big[-\mathbf{S}_1(\mathbf{r})\times\mathbf{S}_5(\mathbf{r}')+\mathbf{S}_1(\mathbf{r})\times\mathbf{S}_6(\mathbf{r}')+\mathbf{S}_2(\mathbf{r})\times\mathbf{S}_4(\mathbf{r}')-\mathbf{S}_2(\mathbf{r})\times\mathbf{S}_6(\mathbf{r}')-\mathbf{S}_3(\mathbf{r})\times\mathbf{S}_4(\mathbf{r}')+\mathbf{S}_3(\mathbf{r})\times\mathbf{S}_5(\mathbf{r}')\Big]\notag\\
    &+ J_4\sum_{\langle\mathbf{r}\mathbf{r}'\rangle}\Big[\mathbf{S}_1(\mathbf{r})\cdot\mathbf{S}_4(\mathbf{r}')+\mathbf{S}_2(\mathbf{r})\cdot\mathbf{S}_5(\mathbf{r}')+\mathbf{S}_3(\mathbf{r})\cdot\mathbf{S}_6(\mathbf{r}')\Big],
    \label{eq:multi Kagome hamiltonian}
\end{align}
\end{widetext} where $p_1$ and $p_2$ correspond to planes with spins $\mathbf{S}_1$, $\mathbf{S}_2$, $\mathbf{S}_3$, and $\mathbf{S}_4$, $\mathbf{S}_5$, $\mathbf{S}_6$ respectively.
As in the previous section, we have assumed perfect kagome planes, which corresponds to setting $J_4=J_5$.
The out-of-plane exchange anisotropy ($A_1$, $A_4$) was ignored for simplicity.
First, we investigate the effects of the interplane exchange, $J_1$, and DM interaction, $D_1$, on the magnitude of the induced magnetic moment.
As shown in Fig.~\ref{fig: m vs J1}, the antiferromagnetic ($J_1>0$) NN interplane coupling reduces the value of $m$, until at $J_1\approx 1.7$ the ground state changes from $\mathbf{A}+\mathbf{M}$ configuration to a structure with magnetic wavevector $\mathbf{Q}=\left(\frac{1}{3},\frac{1}{3},0\right)$.
\noindent
Similarly, for $D_1>0$ and $J_1 = 0.5$, the magnitude of the magnetic moment slowly decreases.
In the absence of the anisotropic interactions, $D_1>0$ stabilizes the $\mathbf{A}$ configuration, and so we find that even when $D_1 = 10 J_1$, the magnitude of the magnetic moment is small but non-zero.
When the interplane coupling is ferromagnetic ($J_1<0$), the in-plane magnetic moments align antiparallel along the $c$-axis such that the total magnetization is zero (Fig.~\ref{fig:GS 2 planes anis}).
\noindent
For $D_1<0$, the orientation of the spins changes to point along the local anisotropy axes.
Note that the ferromagnetic NNN interactions ($J_4<0$), suggested in previous studies~\cite{Cable_PhysRevB.48.6159,Chen_PhysRevB.102.054403,Park2018} do not introduce any additional energetic competitions and hence do not change the spin structure presented above.

\subsection{\label{subsec: elastic scattering}Elastic neutron scattering}

The effects of the anisotropic interactions on the magnetic ground states can be studied with elastic neutron scattering.
In the following, we ignore the effects of the temperature and consider the case of a single magnetic domain.
\begin{figure}[!t]
    \centering
    \includegraphics[width=0.48\textwidth]{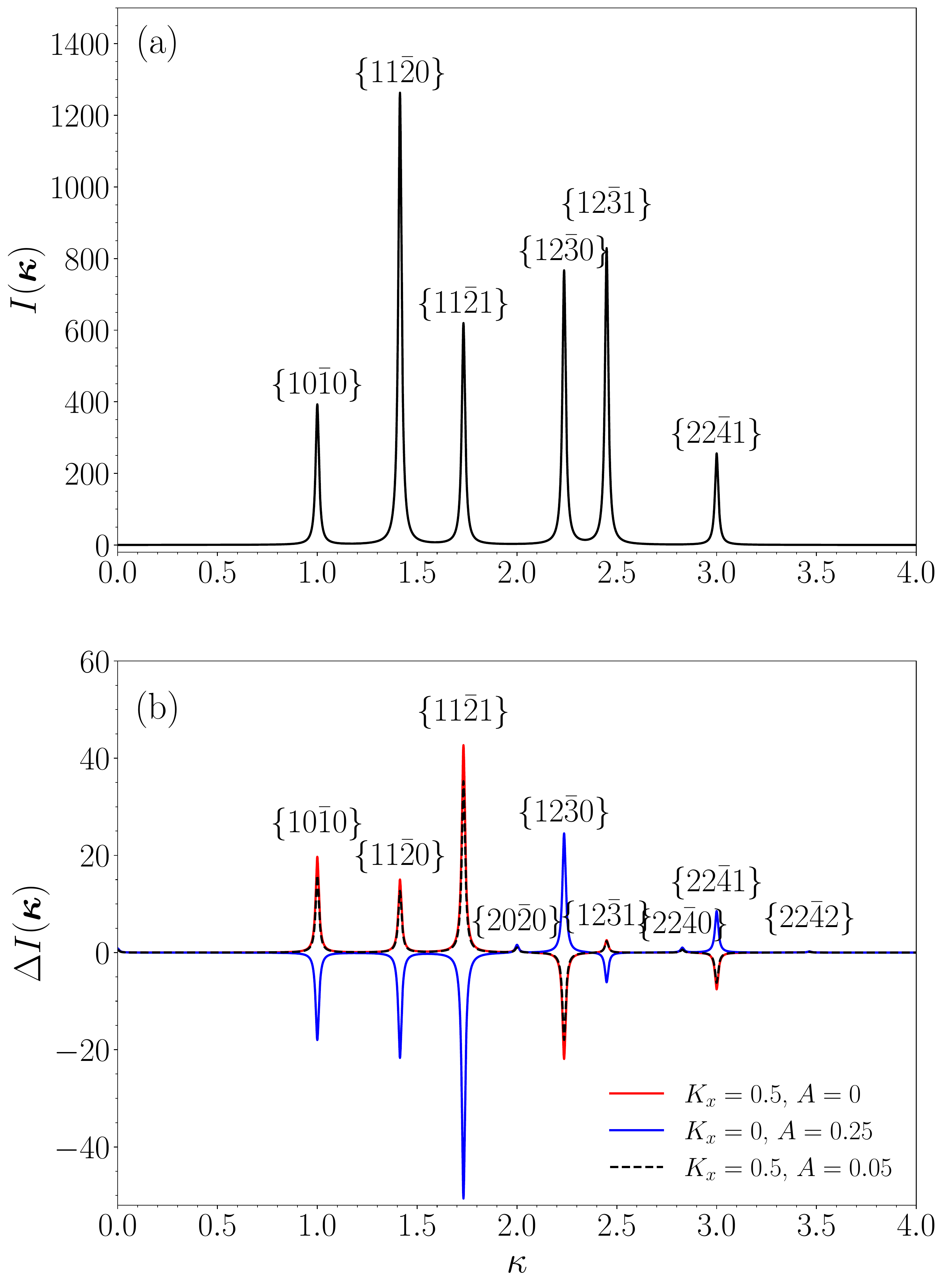}
    \caption{(a) magnetic elastic neutron scattering intensity without anisotropic interactions. (b) effects of the single-ion and exchange anisotropy on the intensity. The intensity difference, $\Delta I(\boldsymbol{\kappa})$ is calculated by subtracting the intensity of the magnetic system without anisotropic interactions. For both panels, $J_2=0.5$ and $D_2=0$ were used.}
    \label{fig:Intensity elastic}
\end{figure}
\noindent
The elastic scattering differential cross-section is proportional to the quantity~\cite{Marshall_NS,squires_2012}
\begin{equation}
    I(\boldsymbol{\kappa}) = |F(\boldsymbol{\kappa})|^2 \sum_{\alpha\beta} \mathcal{S}_{\alpha\beta}(\boldsymbol{\kappa})(\delta_{\alpha\beta} - \hat{\kappa}_\alpha\hat{\kappa}_\beta),
    \label{eq: elastic intensity}
\end{equation} where $\boldsymbol{\kappa}$ is the scattering vector, $F(\boldsymbol{\kappa})$ is the magnetic form factor, obained from the dipolar approximation~\cite{squires_2012,Marshall_NS}, and $\mathcal{S}_{\alpha\beta}(\boldsymbol{\kappa})$ is the static magnetic structure factor:

\begin{equation}
    \mathcal{S}_{\alpha\beta}(\boldsymbol{\kappa}) = \sum_{ij} \langle S_{i\alpha}S_{j\beta}\rangle e^{i\boldsymbol{\kappa}\cdot(\mathbf{r}_i-\mathbf{r}_j)},
\end{equation}

This expression is calculated assuming that the induced magnetic moment points along $\hat{\mathbf{n}}_{4y}$ (same as in Fig.~\ref{fig:GS 2 planes anis}(c) and (d)).
Fig.~\ref{fig:Intensity elastic} shows the effects of the anisotropic interactions on the elastic scattering intensity.
The peaks correspond to a summation over the multiplicity for a given set of $h$, $k$, and $l$ values and therefore would be appropriate for a powder sample.
In the absence of anisotropic interactions, the spectrum displays six peaks dictated by the Bragg reflection conditions.
When the anisotropic terms are included and an in-plane magnetic moment is induced, three additional peaks appear at $\{20\bar{2}0\}$, $\{22\bar{4}0\}$, and $\{22\bar{4}2\}$. 
Note that the intensity of these new peaks is much smaller than the principal peaks.
Nevertheless, it could potentially be enhanced with an applied magnetic field.
Comparing the relative intensity ratios of the principal peaks might allow one to differentiate between the types of magnetic anisotropy in a given material; however, inelastic neutron scattering may give better qualitative signatures of the two types of the anisotropy, as discussed below.

\section{\label{sec:spin_waves}Spin-wave excitations}

We study the impact of the single ion and exchange anisotropies on the spin-wave fluctuations about the ground state configurations by considering the plane-wave solutions of the linearized spin torque equations.
\begin{figure}[h!]
    \centering
    \includegraphics[width=0.45\textwidth,keepaspectratio]{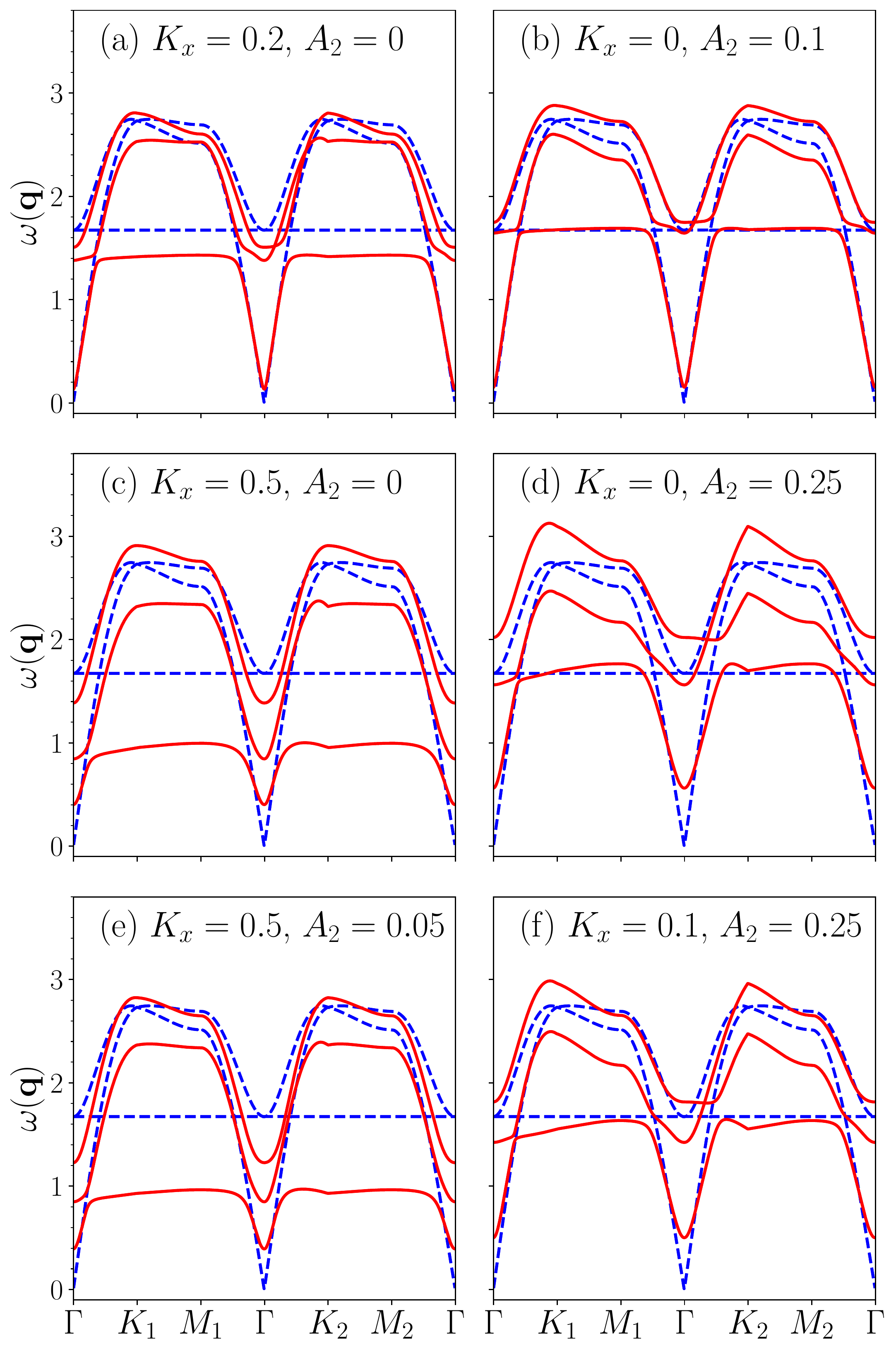}
    \caption{Effects of anisotropic interactions on the spin-wave excitations for a single kagome plane. (a) and (b) represent systems with weak single type of anisotropy; (c) and (d) show systems with strong single type of anisotropy, and (e) and (f) display systems with mixed types of anisotropy (one dominating over another). The dashed blue lines correspond to $K_x=0$, $A_2=0$, and the solid red lines correspond to the dispersion with anisotropic interactions.}
    \label{fig:SW single plane}
\end{figure}
\noindent
In order to perform the spin-wave analysis, it is convenient to introduce a local coordinate system on each sublattice site such that the equilibrium directions of the spins coincide with the $z-$components of the local coordinates~\cite{Leblanc_SW_PhysRevB}.
This is done with the use of six transformation matrices, $\mathbf{U}_i$, which allow one to transform the local spin coordinates into the global ones: $\mathbf{S}_i(\mathbf{r},t) = \mathbf{U}_i\mathbf{\tilde{S}}_i(\mathbf{r},t)$.
Here, the local coordinates are indicated by the tildes.
The collective spin-wave modes are then described with the use of the Fourier transform of the spin density: $\mathbf{\tilde{S}}_i(\mathbf{r},t) = \frac{1}{\sqrt{N}}\sum_\mathbf{q}\mathbf{\tilde{S}}_i(\mathbf{q})e^{i\mathbf{q}\cdot{\mathbf{r}}-i\omega t}$.
The linearized equations of motion simplify to 

\begin{equation}
    -i\omega \tilde{S}_{\alpha i}(\mathbf{q}) = \Gamma_{\alpha\beta ij} \tilde{S}_{\beta j}(\mathbf{q}),
    \label{eq:Equation of motion}
\end{equation} where Einstein summation is implied.
Combining the spin component and sublattice indices, (\ref{eq:Equation of motion}) becomes an eigenvalue problem, which in general must be solved numerically.
In all of the following calculations, the ground state magnetic structures are assumed to have an induced magnetic moment in the $\hat{\mathbf{n}}_{4y}$ direction.

\subsection{\label{subsec: SW one plane} Single layer}

First, we consider the effects of single-ion and exchange anisotropies in a single kagome plane where the magnetic structure is determined by minimizing (\ref{eq:single Kagome hamiltonian}).
The solutions of the linearized equations of motion (\ref{eq:Equation of motion}) correspond to three spin-wave modes. 
When only isotropic exchange interaction are present in the system, all three modes are gapless with a single dispesionless mode, which reflects the macroscopic degeneracy of the 120$^\circ$ ground state~\cite{Harris,Leblanc_SW_PhysRevB,Zhitomirsky_SW_PhysRevB}.
Intraplanar NN DM coupling ($D_2$) lifts the energies of two of the modes resulting in one acoustic and two optical modes~\cite{Zhitomirsky_SW_PhysRevB}. 
The corresponding frequencies are

\begin{align}
    \omega_{1,2}^2 & = 3(J_2-\sqrt{3}D_2)^2- J_2(J_2-\sqrt{3}D_2)f(\mathbf{q})\notag\\
    &\pm D_2(J_2-\sqrt{3}D_2)\sqrt{9+6f(\mathbf{q}))},\\
    \omega_{3}^2 &= 6D_2(3D_2-\sqrt{3}J_2).
\end{align} where 

\begin{equation}
    f(\mathbf{q})=\cos(2q_x)+\cos\left(q_x+\sqrt{3}q_y\right) + \cos\left(q_x-\sqrt{3}q_y\right)
\end{equation} where $q_x$ and $q_y$ lie within the first Brillouin zone and the NN lattice parameter, $a$, was set to 1. 
Note that the DM interactions lift the dispersionless mode to a finite frequency.
Near the $\Gamma$ point, $\mathbf{q}=0$, the expressions for the frequencies become

\begin{align}
    &\omega_1 \approx \sqrt{(J_2-\sqrt{3}D_2)(3J_2-\sqrt{3}D_2)}|\mathbf{q}|\\
    &\omega_{2} \approx \sqrt{6D_2(3D_2-\sqrt{3}J_2)+(J_2-\sqrt{3}D_2)(3J_2+\sqrt{3}D_2)|\mathbf{q}|^2},\\
    &\omega_{3} = \sqrt{6D_2(3D_2-\sqrt{3}J_2)}.
    \label{eq:JDM frequency}
\end{align} Thus, $D_2<0$ enhances the velocity of $\omega_1$ and $\omega_2$ at the $\Gamma$ point.
The spin-wave modes with and without anisotropy are presented in Fig.~\ref{fig:SW single plane}.
When $D_2$ is zero, these expressions simplify to $\omega_{1,2} = \sqrt{3}J_2|\mathbf{q}|$ and $\omega_3 = 0$, which is consistent with Eq.~69 in Ref.~\cite{Dasgupta_PhysRevB.102.144417}.
The inclusion of anisotropic interactions breaks the degeneracy of the doublet and introduces a gap in the dispersion of the acoustic mode.
Here, $K_1 = \Big(\frac{1}{3},\frac{1}{3},0\Big)$, $K_2 = \Big(\frac{2}{3},\bar{\frac{1}{3}},0\Big)$, $M_1 = \Big(\frac{1}{2},0,0\Big)$, $M_2 = \Big(\frac{1}{2},\bar{\frac{1}{2}},0\Big)$.
The different symmetry points were chosen to be at 60 degrees to each other.
An important feature that can be observed throughout these results is that these modes break the six-fold rotational symmetry of the material, which can be seen by comparing the dispersion in the $\Gamma K_1$ and $\Gamma K_2$ regions.
The reason for this symmetry breaking is the fact that the induced magnetic moment pins the ground state configuration with only one of the sublattices oriented parallel to its respective local axis, determined by the anisotropic interactions.
The results in Fig.~\ref{fig:SW single plane} are presented in such a way as to compare the dispersion for both weak and strong single type (either single ion or exchange) anisotropy as well as the modes corresponding to the systems with mixed anisotropic interactions.
It is clear that the qualitative features of the spin-wave dispersion (such as energy gaps) depend strongly on the anisotropic terms even when the corresponding coupling constants are small.
The calculations of the analytical expressions for the energy gaps at the $\Gamma$ point are given in the Supplemental Material.

\subsection{\label{subsec: SW multiple planes} AB-stacked layers planes}

When the AB-stacked kagome planes become coupled, the equations of motion produce six independent modes, three of which are  acoustic and three are optical.
When both $D_1$ and $D_2$ are zero, the acoustic modes are gapless, and the optic modes form a singlet and a doublet.
The velocities of these modes at the $\Gamma$ point depend on the relative strengths of the interplanar coupling constants.
The velocity of the lowest energy mode is given by

\begin{align}
    v_1^{xy} &= a\sqrt{\frac{3(J_2-4J_4)(J_1+J_2)(3J_2+2J_1-12J_4)}{3J_2+J_1-12J_4}},\\
    v_1^{z} &= c\sqrt{6(J_1+J_2)(J_1-3J_4)}.
\end{align} Note that the velocities of the spin-wave modes have been previously calculated in Refs.~\cite{Chen_PhysRevB.102.054403,Dasgupta_PhysRevB.102.144417}, however these references used different notation for the NNN coupling constants: $J_4$ and $J_5$ constants in the present work were labelled as $J_3$ and $J_4$ respectively. 
Furthermore, the references mentioned above set $J_4=0$ ($J_3=0$ in the alternative notation), whereas, as mentioned previously, we have set $J_4=J_5$ throughout the paper.
As a result, the expression for the out-of-plane velocity ($v_1^{(z)}$) presented here is equivalent to the previously reported expressions (Eq.~7 in Ref.~\cite{Chen_PhysRevB.102.054403} and Eq.~73 in Ref.~\cite{Dasgupta_PhysRevB.102.144417}), within the assumptions made in regards to the NNN coupling constants.
However, the expression for the in-plane velocity ($v_1^{(xy)}$) is very different from those presented previously, although it yields similar numerical values.
The origins of this discrepancy are unclear.
The velocities for the remaining modes are given in the Supplemental Material.
DM interactions lift the energy of two of the acoustic modes leading to a gapless singlet and a gapped doublet. 
The corresponding squared frequencies are 
\begin{figure}[!t]
    \centering
    \includegraphics[width=0.49\textwidth,keepaspectratio]{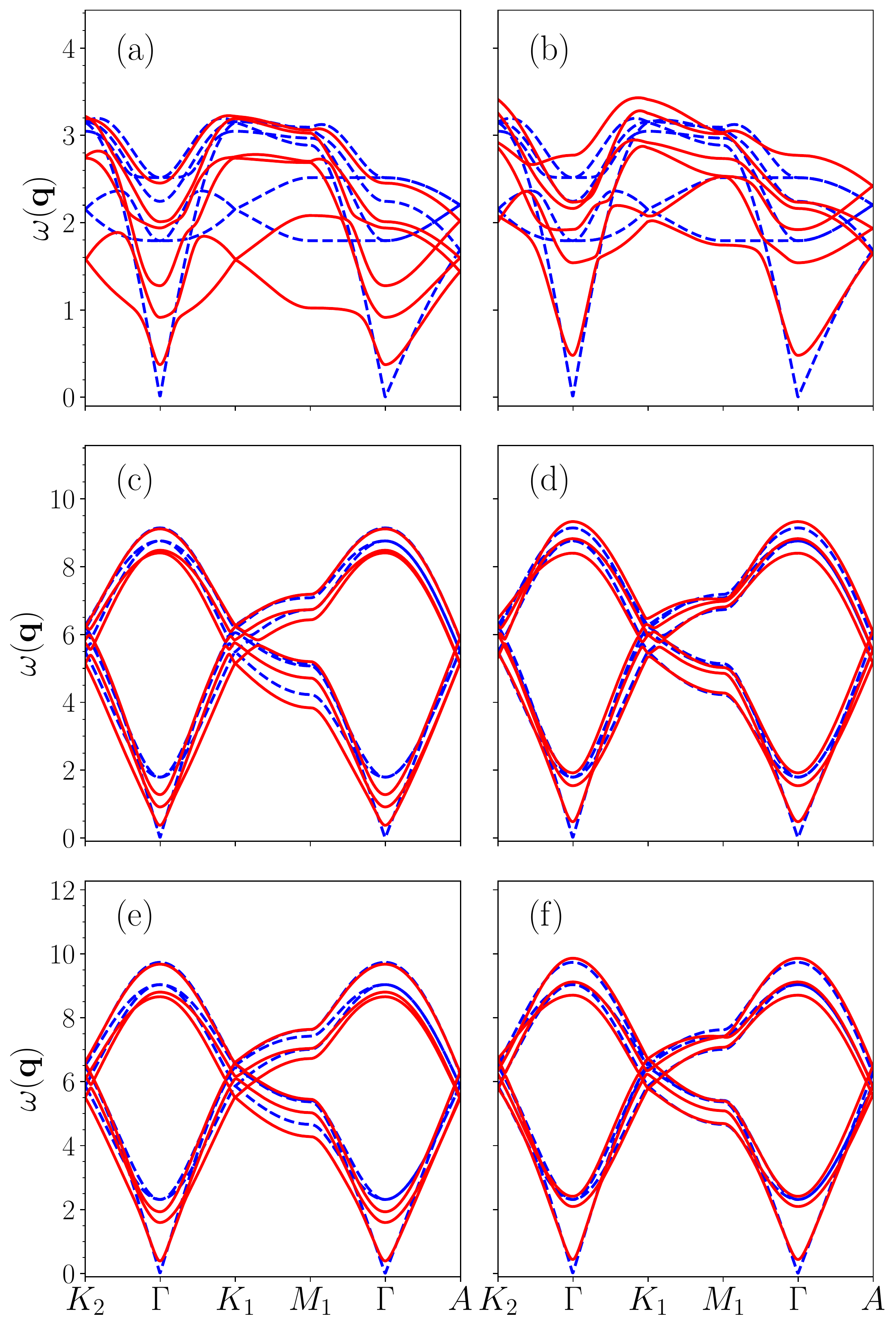}
    \caption{Spin-wave modes in AB-stacked kagome antiferromagnet. Here, $A = \Big(0,0,\frac{1}{2}\Big)$. In (a), (c), (e) the anisotropy parameters are set to $K_x=0.5$, $A_2=0.05$, and in (b), (d), (f) they are $K_x=0.1$, $A_2=0.25$. (a) and (b) show the dispersion of a system of kagome planes coupled via NN exchange ($J_1=0.2$) only, (c) and (d) include also the NNN inter-planar exchange coupling ($J_4=-0.5$), and finally, (e) and (f) also have interplane DM interaction ($D_1=0.1$). The dashed blue lines correspond to $K_x=0$, $A_2=0$, and the solid red lines correspond to the dispersion with anisotropic interactions.}
    \label{fig:SW multiple planes}
\end{figure}
\begin{figure*}[!th]
    \center
    \includegraphics[width=0.94\textwidth]{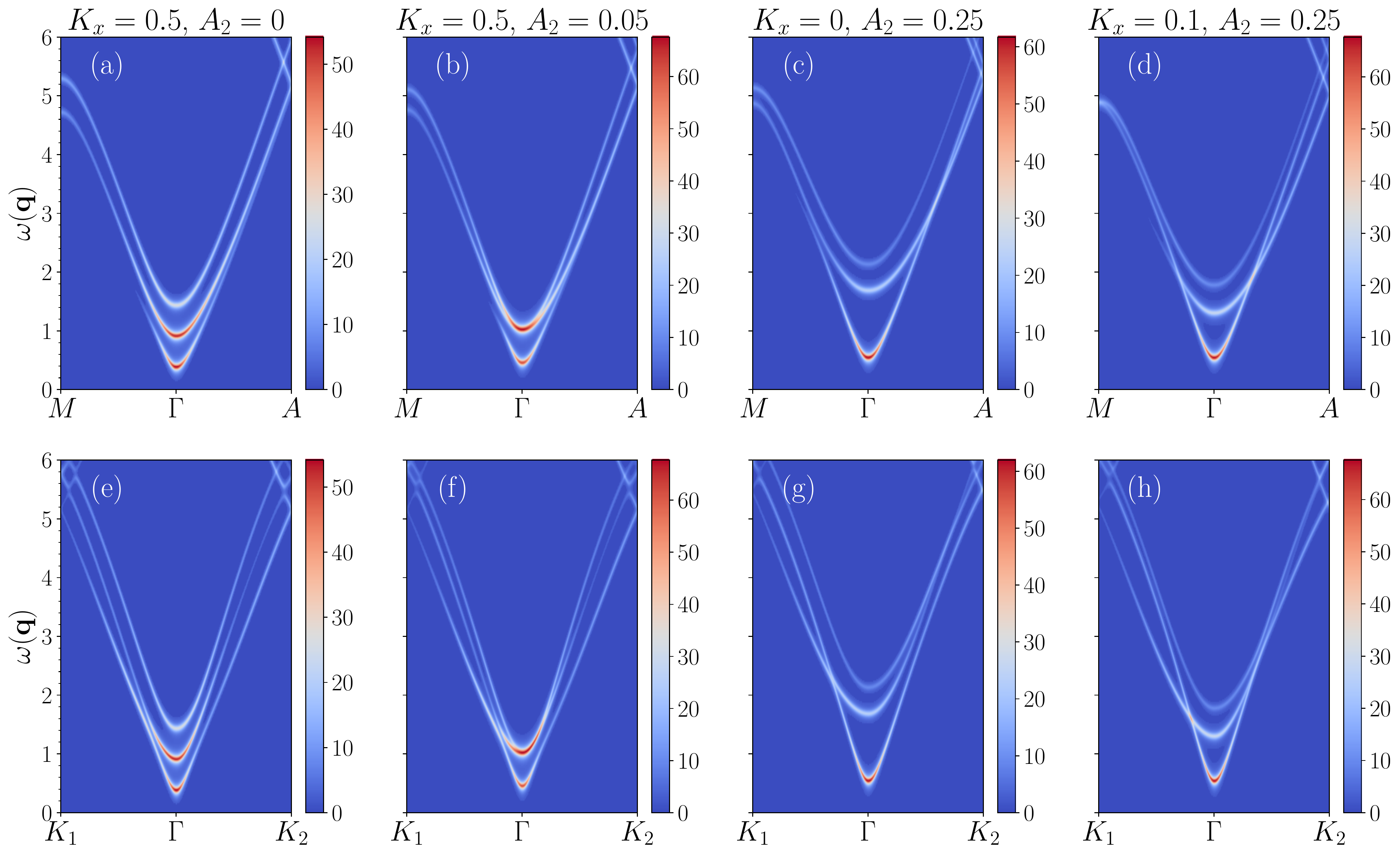}
    \caption{Relative magnitude of the inelastic scattering function $\mathcal{S}(\mathbf{q},\omega)$ (colorbar scale) near the $\Gamma$ point, $\mathbf{q} = [11\bar{1}0]$, assuming that the induced magnetic moment is parallel to $\hat{\mathbf{n}}_{4y}$. (a), (e) and (c), (g) correspond to single kind of anisotropy (single-ion and exchange respectively), (b), (f) and (d), (h) have both kinds of anisotropy. We have set $J_1=0.2$, $J_4 =-0.5$, and $D_1=0$.}
    \label{fig:INS}
\end{figure*}
\begin{align}
    \omega_{1}^2 &= 0,\label{MP_freq_1}\\
    \omega_{2,3}^2 &= 6\sqrt{3}(J_1+J_2)(D_1-D_2)+18(D1-D2)^2\label{MP_freq_23}\\
    \omega_{4,5}^2 &= 6 D_{1}^{2} - 24 D_{1} D_{2} + 6 \sqrt{3} D_{1} J_{1} + 6 \sqrt{3} D_{1} J_{2} + 144 J_{4}^{2}\notag\\
    &- 36 \sqrt{3} D_{1} J_{4} + 18 D_{2}^{2} - 14 \sqrt{3} D_{2} J_{1} - 6 \sqrt{3} D_{2} J_{2} \notag\\
    &+ 60 \sqrt{3} D_{2} J_{4}+ 4 J_{1}^{2}+ 12 J_{1} J_{2} - 60 J_{1} J_{4} - 36 J_{2} J_{4},\\
    \omega_{6}^2 &= 24 D_{1}^{2} - 24 D_{1} D_{2} + 24 \sqrt{3} D_{1} J_{2} - 72 \sqrt{3} D_{1} J_{4}\notag\\
    &- 8 \sqrt{3} D_{2} J_{1}+ 24 \sqrt{3} D_{2} J_{4} - 8 J_{1}^{2} + 24 J_{1} J_{2} - 24 J_{1} J_{4}\notag\\
    &- 72 J_{2} J_{4} + 144 J_{4}^{2}.\label{MP_freq_6}
\end{align}
Frequencies $\omega_{1-3}$ correspond to acoustic modes and $\omega_{4-6}$ -- to optical modes.
Note that the NNN exchange interactions ($J_4$) only enter the expressions~\ref{MP_freq_1}~-~\ref{MP_freq_6} for the optic modes. 
The dispersion of the spin-wave modes for AB-staked kagome planes are presented in Fig.~\ref{fig:SW multiple planes}.
We choose the values of the interplane coupling based on the previous estimates of the exchange parameters for $\mathrm{Mn}_3\mathrm{Ge}$~\cite{Cable_PhysRevB.48.6159,Park2018,Chen_PhysRevB.102.054403}.
Thus, we choose $J_4$ to be large (and negative), and $J_1$ to be small (and positive).
The value of $D_1$ was chosen to be half that of $J_1$.
The two cases considered here correspond to one type of anisotropy dominating over another with the values of $K_x$ and $A_2$ chosen to give similar magnitudes of induced magnetic moment.
In both cases, the strong ferromagnetic NNN exchange ($J_4$) interactions lead to a large energy gap between the three lowest and three highest energy modes.
In an actual experiment, the latter typically appear at very high energies and are often unobservable~\cite{Cable_PhysRevB.48.6159,Radhakrishna_1991,Sukhanov_PhysRevB.99.214445,Chen_PhysRevB.102.054403}. 
The main qualitative differences between the two anisotropy regimes correspond to the branch crossings between the three lowest energy modes.
The small inter-planar DM interaction ($D_1$) moves the second and third lowest energy modes higher in energy and closer together, eventually making them nearly degenerate.
\begin{figure*}[t!]
    \center
    \includegraphics[width=0.94\textwidth]{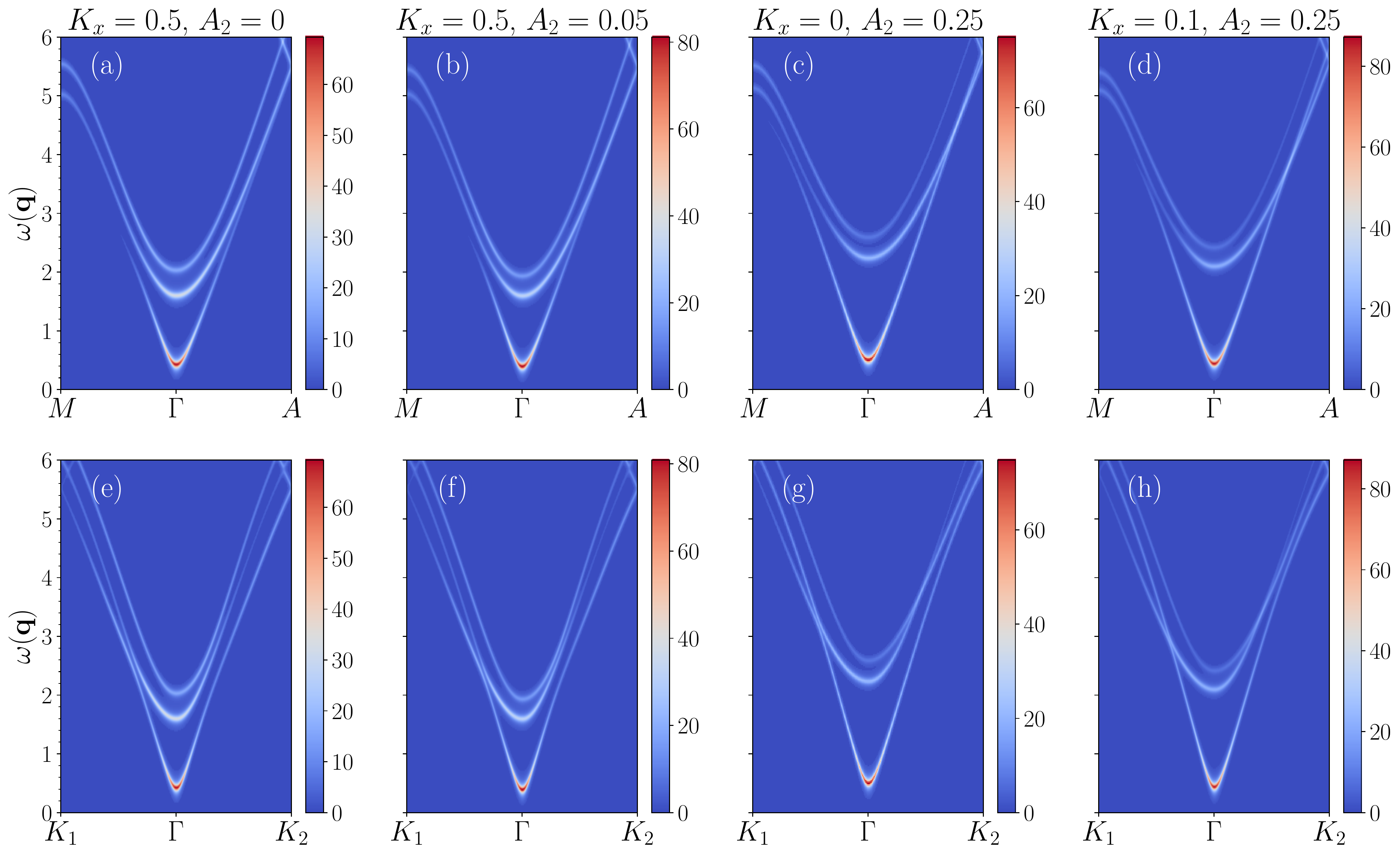}
    \caption{Relative magnitude of dynamic structure factor, $\mathcal{S}(\mathbf{q},\omega)$, (colorbar scale) near $\mathbf{q} = [11\bar{1}0]$ where we have set $J_1=0.2$, $J_4 =-0.5$, and $D_1=0.1$.}
    \label{fig:INS D}
\end{figure*}
\section{\label{sec:inelastic_scattering}Inelastic neutron scattering}

The inelastic scattering cross-section can be written as

\begin{equation}
    \frac{d^2\sigma}{d\Omega dE'} = \left(\frac{\gamma e^2}{m_ec^2}\right)^2\frac{k'}{k}\sum_{\alpha\beta}(\delta_{\alpha\beta}-\hat{q}_\alpha\hat{q}_\beta)\mathcal{S}_{\alpha\beta}(\mathbf{q},\omega),
\end{equation}where 

\begin{equation}
    \mathcal{S}_{\alpha\beta}(\mathbf{q},\omega) = \frac{1}{2\pi\hbar}\int_{-\infty}^\infty dt e^{-i\omega t}\langle T_{\alpha} (\mathbf{q},t) T_{\beta}(-\mathbf{q},t)\rangle
\end{equation} is the dynamic structure factor, and

\begin{equation}
    T_{\alpha}(\mathbf{q},t) = F(q)\sum_{\mathbf{r}}\sum_i\sum_a e^{-i\mathbf{q}\cdot(\mathbf{r}+\mathbf{r}_i)} \Lambda_{\alpha a i}\tilde{S}_{a i}(\mathbf{r},t).
\end{equation} The dynamic structure factor can be calculated using standard Green's function methods~\cite{Leblanc_SW_PhysRevB,Marshall_NS,squires_2012}.
As mentioned earlier, previous inelastic scattering experiments for $\mathrm{Mn}_3\mathrm{Ge}$~\cite{Cable_PhysRevB.48.6159,Radhakrishna_1991,Sukhanov_PhysRevB.99.214445,Chen_PhysRevB.102.054403} and $\mathrm{Mn}_3\mathrm{Sn}$~\cite{Park2018} indicated that the high energy modes are found at $E\sim 100$ meV and are often hard to resolve.
Therefore, we focus on the qualitative features of the lower energy branches.
Fig.~\ref{fig:INS} shows $\mathcal{S}(\mathbf{q},\omega)$ calculated assuming that the induced magnetic moment is oriented along the $\hat{\mathbf{n}}_{4y}$ direction with $J_2 = 1$, $D_2 = -0.2$, $J_1 = 0.2$, $D_1 = 0.1$ and $J_4 = -0.5$.
The intensity is largest for $\mathbf{q}$-vectors which give the elastic peaks, but for smaller wave vectors it drops rapidly; however, the lower energy modes should be distinguishable in an experiment.
Importantly, the characteristic features of the dispersion curves in Fig.~\ref{fig:SW multiple planes} are clearly seen and can be used to deduce the types of anisotropic interactions in the $\mathrm{Mn}_3\mathrm{X}$ compounds.
First of all, the dominant anisotropic interactions can be deduced from the separation of spin-wave branches and the asymmetry of the modes around the $\Gamma$ point.
From there, the possibility of mixed types of anisotropic interactions can be investigated by comparing the relative energy gaps and the velocities of the spin-wave modes, as well as, where possible, the qualitative features like the branch crossings.

Fig.~\ref{fig:INS D} shows the effects of the interplane DM interaction on the relative intensities of the dynamic structure factor.
As stated in the previous section, the second and third lowest energy modes are pushed closer to each other, and could potentially appear as a single line in an experiment.
This situation could be the case for the spin-wave spectra for $\mathrm{Mn}_3\mathrm{Ge}$~\cite{Sukhanov_PhysRevB.99.214445,Chen_PhysRevB.102.054403}, although further experimental studies might be illuminating.

\section{\label{sec:conclusions}Conclusions}

In summary, we have used symmetry considerations in order to construct a general magnetic Hamiltonian for AB-stacked magnetic kagome planes with hexagonal symmetry, with a focus on $\mathrm{Mn}_3\mathrm{X}$ compounds.
In addition to the previously known interactions we have also derived from symmetry an additional NN inter-planar DM coupling, as well as symmetric anisotropic exchange interactions.
The magnetic ground state of the $\mathrm{Mn}_3\mathrm{X}$ systems, which corresponds to the distorted inverse-triangular structure, was shown to depend strongly on the anisotropic terms in the the model.
In particular, the magnitude and direction of the in-plane magnetic moment, induced by the distortion of the 120$^\circ$ state, is determined by the relative strengths of the single-ion and exchange anisotropies.
In either case, the anisotropy pins the ground state removing the continuous degeneracy of the 120$^\circ$ configuration.

Bond-dependent anisotropic exchange interactions in bulk magnetic systems have been, for the most part, neglected in the literature, despite having similar physical origin as the DM interaction.
Nevertheless, as some recent studies indicate, this type of interaction can be crucial for understanding the magnetic properties of some materials. 
In particular, Kitaev-type interactions in honeycomb $\mathrm{Na}_2\mathrm{IrO}_3$ have been shown to dominant over isotropic antiferromagnetic exchange~\cite{Sizyuk,Katukuri_2014}.
Another example is the recently synthesized compound $\mathrm{YbMgGaO}_4$ with triangular lattice structure where the exchange anisotropy was argued to stabilize the quantum spin liquid ground state~\cite{Li_spin_liquid,PhysRevX}.

The two types of the anisotropic interactions have opposing effects on the elastic scattering intensity; however distinguishing between the different kinds of anisotropic interactions from the elastic scattering experiments only may be challenging.
On the contrary, the spin-wave excitations were shown to be very sensitive to even small changes in anisotropic coupling constants which makes the inelastic neutron scattering a better candidate for studying the anisotropic effects in these compounds.
Both kinds of anisotropy break the degeneracy of the optical modes and introduce a gap in the acoustic mode.
The dispersion was also shown to break the six-fold rotational symmetry, reflecting the ``pinning'' of the ground state.
Most importantly, we have shown that the characteristic features of the excitation spectra for systems with one or two kinds of magnetic anisotropy should be accessible in an experimental setting. 

\section{\label{sec: acknowledgements} Acknowledgements}

The authors would like to thank J. S. R. McCoombs and S. H. Curnoe for very valuable discussions and suggestions. This work was supported by the Natural Sciences and Engineering Research Council of Canada (NSERC).
\bibliography{references.bib}

\end{document}